\documentclass[12pt]{article}
\begin{document}
\title{ AVERAGE (RECOMMENDED) HALF-LIFE VALUES FOR TWO NEUTRINO 
DOUBLE BETA DECAY } 
\author{A.S. Barabash \\[0.4cm]
{\small Institute of Theoretical and Experimental Physics, B.\
Cheremushkinskaya 25,} \\ 
{\small 117259 Moscow, Russia}
}
\date{ }
\maketitle
\begin{abstract}
All existing "positive" results on two neutrino double beta decay in 
different nuclei 
were analyzed. Using  procedure recommended by Particle Data Group weighted 
average values for half-lives of $^{48}$Ca, $^{76}$Ge, $^{82}$Se, $^{96}$Zr,
 $^{100}$Mo, $^{100}$Mo - $^{100}$Ru ($0^+_1$), $^{116}$Cd, 
 $^{150}$Nd and $^{238}$U were obtained. 
Existing geochemical data were analyzed 
and recommended values for half-lives of $^{128}$Te and $^{130}$Te  
are proposed.
     We recommend to use these results as most precise and reliable values
 for half-lives at this moment. 
\end{abstract}

\section{Introduction}   

At present two neutrino double beta decay process ($2\nu\beta\beta$
) has been detected in 10 
nuclei. In $^{100}$Mo this type of decay was detected for the transition to $0^+$ 
excited state of daughter nucleus 
too. All these results were obtained in a few tens of geochemical experiments, 
more then two tens of direct (counting) experiments and in one radiochemical 
experiment. In direct experiments for some nuclei there are up to 5 independent 
positive results ($^{76}$Ge, $^{100}$Mo). In some experiments, statistical error 
does not play 
the main role. Thus in Heidelberg-Moscow experiment with $^{76}$Ge more then 
64000 useful events were detected \cite{GUN97,KLA01} and in NEMO experiment with 
$^{100}$Mo 
1423 events were detected \cite{DAS95}. This results in values for the statistical 
error of ~ 
0.5\%  and ~ 4\%, respectively. At the same time systematic error in the 
experiments on $2\nu\beta\beta$
 decay remain quite high ($\sim 10-30\%$) and, in addition, very 
often it can't be determined reliably enough. For example, in Geidelberg-Moscow 
experiment the measured half-life value recently was changed by 15\% with an 
indicated systematic error $\sim 7\%$ (see \cite{GUN97}  and \cite{KLA01}). This  
demonstrates that the real 
systematic error in this experiment \cite{GUN97}  was higher than the
 indicated one. And now authors estimate the systematic error as
  $\sim 10-12\%$ \cite{KLA01}. 

     In the present work critical analysis of all "positive" experimental
results are 
made and averaged, and recommended values for all isotopes are obtained.

%Table 1
\begin{table}
\caption{Present "positive" $2\nu\beta\beta$ decay results. 
N is the number of useful events, S/B is the signal/background ratio.  }
\bigskip
\label{Table1}
%%\begin{tabular*}{\textwidth}{l@{\extracolsep{\fill}}cccc@{   }ccc}
%\begin{tabular*}{\textwidth}{l|cccc|c@{\extracolsep{\fill}}cc}
\begin{tabular}{|c|c|c|c|c|}
\hline
\rule[-2.5mm]{0mm}{6.5mm}
Nucleus & N & $T_{1/2}$, y & S/B & Ref., year \\
%\rule[-1mm]{0mm}{3.5mm} 
\hline
\rule[-2mm]{0mm}{6mm}
$^{48}$Ca & $\sim 100$ & $[4.3^{+2.4}_{-1.1}(stat)\pm 1.4(syst)]\cdot 10^{19}$
  & 1/5 & \cite{BAL96}, 1996 \\
 & 5 & $4.2^{+3.3}_{-1.3}\cdot 10^{19}$ & 5/0 & \cite{BRU00}, 2000 \\
\rule[-4mm]{0mm}{10mm}
 & & {\bf Average value:} $\bf 4.2^{+2.1}_{-1.0} \cdot 10^{19}$ & & \\  
          
\hline
\rule[-2mm]{0mm}{6mm}
$^{76}$Ge & $\sim 4000$ & $(0.9\pm 0.1)\cdot 10^{21}$ & $\sim 1/8$                                                        
& \cite{VAS90}, 1990 \\
& 758 & $1.1^{+0.6}_{-0.3}\cdot 10^{21}$ & $\sim 1/6$ & \cite{MIL91}, 1991 \\
& 132 & $0.93^{+0.2}_{-0.1}\cdot 10^{21}$ & $\sim 4$ & \cite{AVI91}, 1991 \\
& 132 & $1.2^{+0.2}_{-0.1}\cdot 10^{21}$ & $\sim 4$ & \cite{AVI94}, 1994 \\
& $\sim 3000$ & $(1.45\pm 0.15)\cdot 10^{21}$ & $\sim 1.5$ & \cite{MOR99}, 1999 
\\
& 64553 & $[1.55\pm 0.01(stat)^{+0.19}_{-0.15}(syst)]\cdot 10^{21}$ & $\sim 1.5$ 
& \cite{KLA01}, 2001 \\
\rule[-4mm]{0mm}{10mm}
& & {\bf Average value:} $\bf 1.42^{+0.09}_{-0.07} \cdot 10^{21}$ & & \\

\hline
\rule[-2mm]{0mm}{6mm}
$^{82}$Se & 149.1 & $[0.83 \pm 0.10(stat) \pm 0.07(syst)]\cdot 10^{20}$ & 2.3 & 
\cite{ARN98}, 1998 \\
& 89.6 & $1.08^{+0.26}_{-0.06}\cdot 10^{20}$ & $\sim 8$ & \cite{ELL92}, 1992 \\
& & $(1.3\pm 0.05)\cdot 10^{20}$ (geochem.) & & \cite{KIR86}, 1986 \\
\rule[-4mm]{0mm}{10mm}
& & {\bf Average value:} $\bf (0.9\pm 0.1)\cdot 10^{20}$ & & \\
 
\hline
\rule[-2mm]{0mm}{6mm}
$^{96}$Zr & 26.7 & $[2.1^{+0.8}_{-0.4}(stat) \pm 0.2(syst)]\cdot 10^{19}$ & 1.9 
& \cite{ARN99}, 1999 \\
& & $(3.9\pm 0.9)\cdot 10^{19}$ (geochem.)& & \cite{KAW93}, 1993 \\
\rule[-4mm]{0mm}{10mm}
& & {\bf Recommended value:} $\bf 2.1^{+0.8}_{-0.4}\cdot 10^{19}$ & & \\

\hline
\rule[-2mm]{0mm}{6mm}
$^{100}$Mo & $\sim 500$ & $11.5^{+3.0}_{-2.0}\cdot 10^{18}$ & 1/7 & 
\cite{EJI91}, 1991 \\
& 67 & $11.6^{+3.4}_{-0.8}\cdot 10^{18}$ & 1 & \cite{ELL91}, 1991 \\
& 1433 & $[9.5 \pm 0.4(stat) \pm 0.9(syst)]\cdot 10^{18}$ & 3 & 
\cite{DAS95}, 1995 \\
& 175 & $7.6^{+2.2}_{-1.4}\cdot 10^{18}$ & 1/2 & \cite{ALS97}, 1997 \\
& 377 & $[6.75^{+0.37}_{-0.42}(stat) \pm 0.68(syst)]\cdot 10^{18}$ & 10 & 
\cite{DES97}, 1997 \\
& 800 & $[7.2 \pm 1.1(stat) \pm 1.8(syst)]\cdot 10^{18}$ & 1/9 & 
\cite{ASH01}, 2001 \\
\rule[-4mm]{0mm}{10mm}
& & {\bf Average value:} $\bf (8.0\pm 0.7)\cdot 10^{18}$ & & \\

\hline
%\rule[-2mm]{0mm}{6mm}
$^{100}$Mo - & 66 & $6.1^{+1.8}_{-1.1}\cdot 10^{20}$ & 1/7 & 
\cite{BAR95}, 1995 \\
$^{100}$Ru ($0^+_1$) & $\sim 80$ & $[9.3^{+2.8}_{-1.7}(stat) \pm 1.4(syst)]\cdot 
10^{20}$ & 1/4 & \cite{BAR99}, 1999 \\
 & 19.5 & $[5.9^{+1.7}_{-1.1}(stat) \pm 0.6(syst)]\cdot 10^{20}$ & $\sim 8$ & 
\cite{DEB01}, 2001 \\   
\rule[-4mm]{0mm}{10mm}
& & {\bf Average value:} $\bf (6.8\pm 1.2)\cdot 10^{20}$ & & \\

\hline
%\rule[-2mm]{0mm}{6mm}
$^{116}$Cd& $\sim 180$ & $2.6^{+0.9}_{-0.5}\cdot 10^{19}$ & $\sim 1/4$ & 
\cite{EJI95}, 1995 \\
& 3600 & $[2.6\pm 0.1(stat)^{+0.7}_{-0.4}(syst)]\cdot 10^{19}$ & $\sim 2$ & 
\cite{DAN00}, 2000 \\
& 174.6 & $[3.75 \pm 0.35(stat) \pm 0.21(syst)]\cdot 10^{19}$ & 3 & 
\cite{ARN96}, 1996 \\
\rule[-4mm]{0mm}{10mm}
& & {\bf Average value:} $\bf 3.3^{+0.4}_{-0.3}\cdot 10^{19}$ & & \\

\hline
\end{tabular}
\end{table}

%Table 1 continued
\addtocounter{table}{-1}
\begin{table}
\caption{continued.}
\bigskip
%\label{Table1}
\begin{tabular}{|c|c|c|c|c|}
\hline
\rule[-2mm]{0mm}{6mm}
$^{128}$Te& & $2.2\cdot 10^{24}$ (geochem.) & & \cite{MAN91}, 1991 \\
& & $(7.7\pm 0.4)\cdot 10^{24}$ (geochem.)& & \cite{BER93}, 1993 \\
\rule[-4mm]{0mm}{10mm}
& & {\bf Recommended value:} $\bf (2.5\pm 0.4)\cdot 10^{24}$ & & \\

\hline
\rule[-2mm]{0mm}{6mm}
$^{130}$Te& & $0.8\cdot 10^{21}$ (geochem.) & & \cite{MAN91}, 1991 \\
& & $(2.7\pm 0.1)\cdot 10^{21}$ (geochem.)& & \cite{BER93}, 1993 \\
\rule[-4mm]{0mm}{10mm}
& & {\bf Recommended value:} $\bf (0.9\pm 0.15)\cdot 10^{21}$ & & \\

\hline
\rule[-2mm]{0mm}{6mm}
$^{150}$Nd& 23 & $[1.88^{+0.69}_{-0.39}(stat) \pm 0.19(syst)]\cdot 10^{19}$ & 
1.8 & \cite{ART95}, 1995 \\
& 414 & $[6.75^{+0.37}_{-0.42}(stat) \pm 0.68(syst)]\cdot 10^{18}$ & 6 & 
\cite{DES97}, 1997 \\
\rule[-4mm]{0mm}{10mm}
& & {\bf Average value:} $\bf(7.0\pm 1.7)\cdot 10^{18}$ & & \\

\hline
\rule[-2mm]{0mm}{6mm}
$^{238}$U& & $\bf (2.0\pm 0.6)\cdot 10^{21}$ & & \cite{TUR91}, 1991 \\
 
\hline
\end{tabular}
\end{table}

\section{ Present experimental data }
Experimental results on $2\nu\beta\beta$
 decay in different nuclei are presented in Table 1. 
For direct experiments number of useful events and signal/background ratio are 
presented.

\section{ Data analysis }
To obtain an average of the data a standard weighted least-squares procedure recommended by
 Particle Data Group \cite{EUR00} was used. Weighted average and error were 
calculated as:
\begin{equation}
\bar x\pm \delta \bar x = \sum w_ix_i/\sum w_i \pm (\sum w_i)^{-1/2} , 
\end{equation} 

   where      $w_i = 1/(\delta x_i)^2$.

Here $x_i$  and $\delta x_i$ are the value and error reported by the i-th 
experiment, and the sums 
run over the N experiments. We then calculate $\chi^2 = \sum w_i(\bar x - x_i)^2$ and 
compare it with 
N - 1, which is the expectation value of  $\chi^2$ if the measurements are from 
a Gaussian 
distribution.

     If  $\chi^2/(N - 1)$ is less than or equal to 1, and there are no known 
problems with the 
data, we accept the results.

     If $\chi^2/(N - 1)$ is very large, we may choose not to use the average at 
all. 
Alternatively, we may quote the calculated average, but then make an educated 
guess 
of the error, a conservative estimate designed to take into account known 
problems 
with the data.

     Finally, if $\chi^2/(N - 1)$ is greater than 1, but not greatly so, we 
still average the data, 
but  we increase our quoted error, $\delta \bar x$ in Eq. (1), by a scale factor S 
defined as 
\begin{equation}
                                 S = [\chi^2/(N - 1)]^{1/2}  ,                                              
\end{equation} 
For averages we add the statistical and systematic errors in quadrature and use 
this 
combined error as $\delta x_i$. 

\noindent    
\underline{3.1. $^{48}$Ca}. There are two independent experiments in which 
$2\nu\beta\beta$
 decay of $^{48}$Ca was 
observed \cite{BAL96,BRU00}. Results are in good agreement, 
but errors are quite large. The weighted average value is:
$$
                           T_{1/2} = 4.2^{+2.1}_{-1.0} \cdot 10^{19} y. 
$$ 

\noindent 
\underline{3.2. $^{76}$Ge}. Let us consider the results of five experiments. 
But, first of all, a few 
additional comments are necessary:

1) Recently the result of the Heidelberg-Moscow group was corrected. Instead of 
the previously published value $T_{1/2} = [1.77\pm 0.01(stat)^{+0.13}_{-
0.11}(syst)]\cdot 10^{21}$ y
 \cite{GUN97}, a new 
value $T_{1/2} = [1.55\pm 0.01(stat)^{+0.19}_{-0.15}(syst)]\cdot 10^{21}$ y
 \cite{KLA01} has been presented. This last value 
has been used in our analysis.

2) In ref. \cite{AVI91} the value $T_{1/2} = 0.92^{+0.07}_{-0.04}\cdot 10^{21}$ 
y
was presented. However, after more 
careful analysis, it was changed to a value of 
$T_{1/2} = 1.2^{+0.2}_{-0.1}\cdot 10^{21}$ y \cite{AVI94}, 
which was used in 
our analysis.

3) The result of work \cite{VAS90} does not agree with more later and more 
precise experiments 
\cite{KLA01,MOR99}. Error presented in \cite{VAS90} looks too small, especially 
taking into account that 
signal/background ratio in this experiment is equal to $\sim 1/10$. Before it was 
mentioned 
\cite{BAR90}, that half-life value in this work can be $\sim 1.5-2$ times higher 
because the 
thickness of dead layer in using Ge(Li) detectors can be different in crystals 
made of 
natural and enriched Ge. Under nonuniformity of external 
background it 
can have an  appreciable influence  on the final result.

     Finally, in calculating the average, only result of experiments with
signal/background
ratio greater then 1 were used, i.e. the results of \cite{KLA01,AVI94,MOR99}. The 
weighted average value is:
$$
    T_{1/2} = 1.43^{+0.09}_{-0.07} \cdot 10^{21} y.
$$ 

\noindent 
\underline {3.3. $^{82}$Se}. There are two independent counting 
experiments and a lot of geochemical measurements $(\sim 20)$. 
Geochemical data are not in good agreement with each other and with 
direct experiments data. Formally the accuracy of geochemical 
measurements can be on the level of a few percent and even better. 
Nevertheless, now the possibility of 
existing large systematic error cannot be excluded (see discussion in 
\cite{MAN86}). It is mentioned in ref. \cite{BAR00} that, if weak 
interaction constant $G_F$ depends of time, then half-life values 
obtained  in geochemical experiments will depend of the  age of the samples. 
This is why to obtain a "present" half-life value of $^{82}$Se, only results 
of direct measurements,  \cite{ARN98} and \cite{ELL92}, were used. 
Result of ref. \cite{ELL87} is the preliminary 
result of \cite{ELL92} and we do not use it in our analysis. 
Notice that "low" error in \cite{ELL92} looks too small. It 
is even smaller than the statistical error! This is why we use here a more
realistic value of 15\%  as an estimation of this error. 
As a result the weighted average value is:
$$
T_{1/2} = (0.9 \pm 0.1) \cdot 10^{20} y.
$$ 

\noindent 
\underline{3.4. $^{96}$Zr}. There are two "positive" results:  
geochemical result \cite{KAW93} and 
result of direct NEMO experiment \cite{ARN99}. Taking into account the comment 
in section 
3.3 we take the value from \cite{ARN99} as "present" half-life value for 
$^{96}$Zr: 
$$
T_{1/2} = 2.1^{+0.8}_{-0.4} \cdot 10^{19} y.                    
$$ 

\noindent 
\underline {3.5. $^{100}$Mo}. Formally there are 6 positive results 
\footnote{We do not consider result of \cite {VAS90a} because possible background contribution to the "effect" was not excluded in this experiment.} . But we do 
not consider preliminary 
result of M.Moe et al. \cite{ELL91} and use their final result \cite{DES97}. The 
following weighted 
average value for half-life is obtained:
$$
T_{1/2} = (8 \pm 0.7)\cdot 10^{18} y .                                   
$$

\noindent 
\underline{3.6. $^{100}$Mo - $^{100}$Ru ($0^+_1$; 1130.29 keV)}. Transition to 
$0^+$ excited state in $^{100}$Ru was 
detected in three independent experiments. Results are in good agreement. 
Weighted 
average value for half-life is:
$$
T_{1/2} = (6.8 \pm 1.2)\cdot 10^{20} y .
$$                                   

\noindent 
\underline{3.7. $^{116}$Cd}. There are three independent "positive" results 
which are in a good 
agreement with each other taking into account the error bars. The waighted average 
value is:
$$          
T_{1/2} = 3.3^{+0.4}_{-0.3} \cdot 10^{19} y.
$$ 

\noindent 
\underline{3.8. $^{128}$Te and $^{130}$Te}. There are only geochemical data for 
these isotopes. Though the half-
life ratio for this isotopes is obtained with good accuracy $(\sim 3\%)$ 
\cite{BER93}, absolute 
values for $T_{1/2}$ are different from one experiment to the next. One group of 
authors \cite{MAN91,TAK66,TAK96} 
gives $T_{1/2} \approx 0.8\cdot 10^{21}$ y  for $^{130}$Te and 
$T_{1/2} \approx  2\cdot 10^{24}$ y
 for $^{128}$Te and another group \cite{KIR86,BER93} - 
$T_{1/2} \approx (2.5-2.7)\cdot 10^{21}$ y and  $T_{1/2} \approx 7.7\cdot 10^{24}$ y, respectively. 
And besides, as a rule, experiments with 
"young" samples ($\sim 100$ million. years) give for half-life of 
$^{130}$Te values $\sim (0.7-0.9)\cdot 10^{21}$ y and 
for "old" samples ($> 1$ billion years)  $\sim (2.5-2.7)\cdot 10^{21}$ y. 
It was even 
assumed that the difference in half-life values could be connected with 
a variation of the week 
interaction constant $G_F$ with time \cite{BAR00}.

     We will estimate the absolute values of $T_{1/2}$ for $^{130}$Te 
and $^{128}$Te using only very 
well known ratios from geochemical measurements and "present" 
half-life value of $^{82}$Se (see section  3.3). First ratio is 
$T_{1/2}(^{130}{\rm Te})/T_{1/2}(^{128}{\rm Te}) = 
(3.52 \pm 0.11)\cdot 10^{-4}$ \cite{BER93}. Second 
ratio is $T_{1/2}(^{130}{\rm Te})/T_{1/2}(^{82}{\rm Se}) = 9.9 \pm 0.6$. 
This value is 
weighted average value from 
three experiments: $7.3 \pm 0.9$ \cite{LIN86}, $12.5 \pm 0.9$ \cite{KIR86} and 
$10 \pm 2$ \cite{SRI73}. It is significant that 
the gas retention age problem has no effect on the half-life ratios. Now using 
"present" $^{82}$Se half-life value $T_{1/2} = (0.9 \pm 0.1)\cdot 10^{20}$ y. 
and value $9.9 \pm 0.6$ for 
$T_{1/2}(^{130}{\rm Te})/T_{1/2}(^{82}{\rm Se})$ ratio one can obtain the half-life value 
for $^{130}$Te:
$$          
T_{1/2} = (0.9 \pm 0.15)\cdot 10^{21} y.
$$ 

 Using $T_{1/2}(^{130}{\rm Te})/T_{1/2}(^{128}{\rm Te}) = 
(3.52 \pm 0.11)\cdot 10^{-4}$ 
\cite{BER93} one can obtain the half-life 
value for $^{128}$Te:
$$          
T_{1/2} = (2.5 \pm 0.4)\cdot 10^{24} y.
$$ 

\underline{3.9. $^{150}$Nd.} The half-life value was obtained in two independent 
experiments, \cite{ART95} and 
\cite{DES97}. But the two results are not in good agreement. Using the relation (1) one 
can obtain $T_{1/2} 
= (7.0 \pm 0.8)\cdot 10^{18}$ y. Taking into account that $\chi^2 > 1$  and S = 
2.2 (see relation (2)) we 
finally obtain:
$$
T_{1/2} = (7.0 \pm 1.7)\cdot 10^{18} y.
$$ 

\underline{3.10. $^{238}$U}.  There is only one positive result from 
radiochemical experiment \cite{TUR91}:
$$          
T_{1/2} = (2.0 \pm 0.6)\cdot 10^{21} y.
$$ 

\section{Conclusion}

Hence all "positive" $2\nu\beta\beta$ results were analyzed and average values 
for half-lives 
were calculated. For $^{128}$Te and $^{130}$Te so-called "recommended" values 
were proposed. 
We recommend the use of exactly these values as most precise and reliable by this 
moment.

\end{document}